\documentstyle[prl,aps,epsfig,amsmath,amsfonts,amssymb,multicol]{revtex}
\begin{document}
\draft
\title{Evidence for Thermal Equilibration in Multifragmentation\\
Reactions probed with Bremsstrahlung Photons}
\author{D.G. d'Enterria,~$^{1,5,}$\cite{subatech}
L.~Aphecetche,~$^{1,}$\cite{subatech}
A.~Chbihi,~$^{1}$
H.~Delagrange,~$^{1,}$\cite{subatech}
J.~D\'{\i}az,~$^{2}$
M.J.~van~Goethem, $^{3,}$\cite{msu}
M.~Hoefman,~$^{3}$
A.~Kugler, $^{4}$
H.~L\"ohner,~$^{3}$
G.~Mart\'{\i}nez, $^{1,}$\cite{subatech}
M.J.~Mora, $^{1,}$\cite{subatech}
R.~Ortega,~$^{5}$
R.~Ostendorf,~$^{3}$
S.~Schadmand,~$^{6}$
Y.~Schutz,~$^{1,}$\cite{subatech}
R.H.~Siemssen, $^{3}$
D.~Stracener,~$^{7}$
P.~Tlusty,~$^{4}$
R.~Turrisi, $^{1,}$\cite{padova}
M.~Volkerts,~$^{3}$
V.~Wagner,~$^{4}$
H.~Wilschut,~$^{3}$ and
N.~Yahlali~$^{2}$
}
\address{$^1$Grand Acc\'el\'erateur National d'Ions Lourds, BP~5027, 14076 Caen Cedex 5, France}
\address{$^2$Institut de F\'{\i}sica Corpuscular, Universitat de Val\`encia-CSIC, Dr. Moliner 50, 46100 Burjassot, Spain}
\address{$^3$Kernfysisch Versneller Instituut, 9747 AA Groningen, The Netherlands}
\address{$^4$Institute of Nuclear Physics, 25068 \v Re\v z, Czech Republic}
\address{$^5$Grup de F\'{\i}sica de les Radiacions, Universitat Aut\`onoma de Barcelona,
08193 Cerdanyola del Vall\`es, Catalonia}
\address{$^6$II. Physikalisches Institut, Universit\"at Gie\ss en, D-35392 Gie\ss en, Germany}
\address{$^7$Oak Ridge National Laboratory, Oak Ridge, Tennessee 37831}
\date{\today}
\maketitle
\begin{abstract}
The production of nuclear bremsstrahlung photons (E$_{\gamma}>$ 30
MeV) has been studied in inclusive and exclusive measurements in
four heavy-ion reactions at 60{\it A} MeV. The measured photon
spectra, angular distributions and multiplicities indicate that a
significant part of the hard-photons are emitted in secondary
nucleon-nucleon collisions from a thermally equilibrated system.
The observation of the thermal component in multi-fragment
$^{36}$Ar\/+\/$^{197}$Au reactions suggests that the breakup of the
thermalized source produced in this system occurs on a rather long
time-scale.
\end{abstract}

\begin{multicols}{2}
Nucleus-nucleus collisions at intermediate energies (20{\it A}
MeV$\leq$$\epsilon_{lab}$$\leq$ 100{\it A} MeV) aim at the study of the
phase diagram of nuclear matter at densities and temperatures where
a transition from the Fermi liquid ground-state to the nucleon gas
phase has been predicted~\cite{enterria:Poch97}. The experimentally
observed production of several intermediate-mass fragments, IMF,
with $3\leq Z\leq 20$ (``nuclear multifragmentation") has usually
been interpreted as a signal of such a liquid-gas phase
transition~\cite{enterria:Poch97,enterria:Gupt00}. One of the key
issues in the field is whether or not an equilibrated system is
created in the course of the
collision~\cite{enterria:Rich00,enterria:Hirs99} and, implicitly,
whether its lifetime is long enough for equilibration to occur.
Despite their low production rates, ``elementary'' particles such
as photons, dileptons and mesons, are unique probes of the
phase-space evolution of nucleus-nucleus collisions
\cite{enterria:Cass90}. At variance with charged particles and
nuclear fragments, photons are primordial observables because they
do not suffer final-state (neither Coulomb nor strong) interactions
with the surrounding medium, thus providing an unperturbed image of
the emission source. Above E$_{\gamma}$ = 30 MeV, many experimental
facts supported by model calculations
\cite{enterria:Cass90,enterria:Nife90,enterria:Schu97} indicate
that photons are mainly emitted during the first instants of the
reaction in incoherent proton-neutron bremsstrahlung collisions,
$pn\rightarrow pn\gamma$, occurring within the participant zone.
Hard-photons have thus been exploited 
to probe the preequilibrium conditions prevailing in the initial
high-density phase of the reaction
\cite{enterria:Schu97,enterria:Mart94}. Recent experimental results
have pointed out an additional source of hard-photon emission,
softer than the one originating in first-chance $pn\gamma$,
accounting for up to a third of the total hard-photon yield
\cite{enterria:Mart95,enterria:Marq95}. The origin of this second
component has been localized in secondary $NN\gamma$ collisions
within a thermalizing system 
\cite{enterria:Schu97,enterria:Mart95,enterria:Marq95}. Such
thermal hard-photons hence constitute a novel and clean probe of
the intermediate dissipative stages of the reaction where nuclear
fragmentation is supposed to take place, characterizing both its
time-scale and the thermodynamical state 
of the fragmenting source(s). This information is
essential to establish whether the physical mechanism driving
multifragmentation is of fast
statistical~\cite{enterria:Bond95,enterria:Gros97},
sequential~\cite{enterria:Frie90} or purely
dynamical~\cite{enterria:Aich91} origin. 
In this Letter, we first establish the thermal origin of
second-chance hard-photons from inclusive measurements in four
heavy-ion reactions at 60{\it A} MeV. Additionally, the concomitant
observation of thermal hard-photon and IMF emission in the
$^{36}$Ar\/+\/$^{197}$Au system indicates that multifragmentation
is, at least for this reaction, a slow process preceded by the
thermalization of the system.

\begin{figure}
 \begin{center}
  \mbox{\epsfxsize=8.5cm \epsfbox{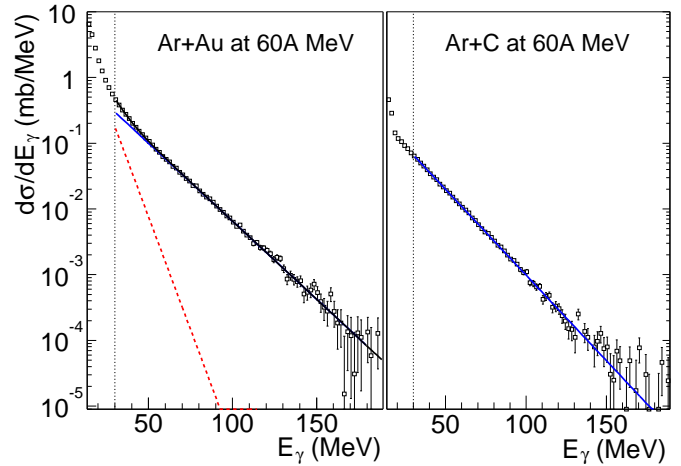}}
 \end{center}
 \caption{Hard-photon (E$_{\gamma}>$30 MeV) spectra for:
$^{36}$Ar\/+\/$^{197}$Au (left), fitted to the sum of a direct
(solid) and a thermal (dashed) exponential distribution; and
$^{36}$Ar\/+\/$^{12}$C (right) fitted to a single exponential.}
 \label{fig:enterria:1}
\end{figure}

\begin{figure}[htbp]
\begin{center}
  \mbox{\epsfxsize=8.5cm \epsfbox{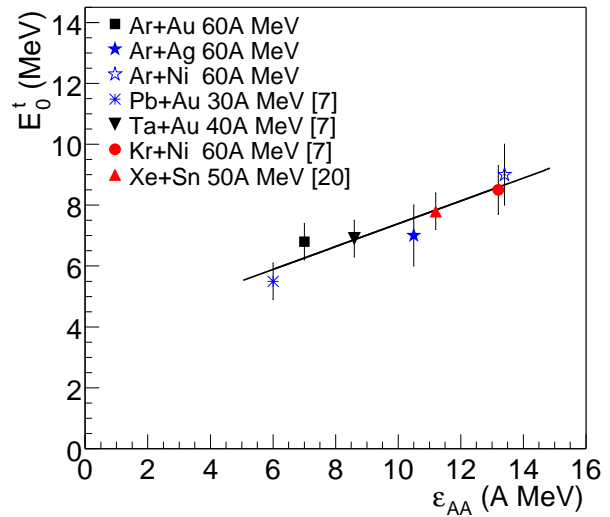}}
 \end{center}
 \caption{Thermal hard-photon slopes $E_0^t$ (measured at $\theta_{lab}$=90$^\circ$)
 as a function of the (Coulomb-corrected) nucleus-nucleus center-of-mass energy, $\epsilon_{AA}$,
 for the 7 reactions studied by TAPS: the 3 heavy systems
 reported here, and those of refs. \protect\cite{enterria:Schu97,enterria:Orte00}.
 The solid line is a linear fit to the data.}
  \label{fig:enterria:2}
\end{figure}

The inclusive and exclusive hard-photon and fragment production in
$^{36}$Ar\/+\/$^{197}$Au at 60{\it A} MeV, and the inclusive
hard-photon production in the $^{36}$Ar\/+\/$^{107}$Ag, $^{58}$Ni,
$^{12}$C reactions at 60{\it A} MeV have been studied at the KVI
laboratory. The $^{36}$Ar beam was delivered by the K = 600 MeV
superconducting AGOR cyclotron with a bunch rate of 37.1 MHz and
with intensities ranging from 3.0 to 12.5 nA. Targets consisted of
thin isotopically enriched foils of 1 to 18 mg/cm$^2$.
The TAPS~\cite{enterria:Novo91} electromagnetic spectrometer,
comprising 384 BaF$_2$ scintillation modules in a six-block
configuration and covering the polar angles between 57$^\circ$ and
176$^\circ$ and the azimuthal range from -20$^\circ$ to 20$^\circ$,
was used to measure the double differential cross section
$d^2\sigma/d\Omega_\gamma dE_\gamma$ for photons of 10 MeV$\leq
E_{\gamma} \leq$ 300 MeV. Photons have been discriminated against
charged particles and neutrons on the basis of pulse-shape
analysis, time-of-flight and charged-particle veto
information~\cite{enterria:Marq95b,enterria:Dente00b}. Two phoswich
multidetectors, the Washington University ``Dwarf-Ball"
(DB)~\cite{enterria:Stra90} and the KVI ``Forward Wall" (FW)
\cite{enterria:Leeg92}, were added to TAPS to allow the isotopic
identification of the light charged particles (LCP: $p$, $d$, $t$,
$^3$He and $\alpha$) and the charge of the IMFs up to that of the
projectile by means of pulse-shape techniques. The DB was composed
of 64 BC400-CsI(Tl) phoswich telescopes in the angular range
$32^\circ < \theta < 168^\circ$ covering about 76\% of 4$\pi$, and
the FW hodoscope comprised 92 NE102A-NE115 $\Delta$E-E phoswich
detectors in the forward region ($2.5^\circ < \theta < 25^\circ$).
Altogether, more than 1.2$\cdot 10^6$ hard-photons were collected
in the four reactions with a trigger defined by at least one
neutral hit in TAPS depositing more than 10 MeV in a single BaF$_2$
crystal, plus three or more hits in the charged particle detectors.
The inclusive photon energy spectra have been obtained after
correction for the detector response and after subtraction of the
cosmic and $\pi^0$-decay (measured via $\gamma\gamma$ invariant
mass analysis) backgrounds which amount to (5$\pm$1)\% of the total
yield \cite{enterria:Dente00b}. The hard-photon spectra of the Au,
Ag and Ni targets feature two distinct components above 30 MeV
(Fig. \ref{fig:enterria:1} left, for the Au target) and can be
described by the sum of two exponential distributions characterized
by inverse slopes $E_0^d$ and $E_0^t$, corresponding to a ``direct"
(first-chance) and a ``thermal" (secondary $pn\gamma$) component,
with their corresponding weights
\cite{enterria:Schu97,enterria:Mart95}:
\begin{equation}
\label{eq:enterria:1}
\frac{d\sigma}{d E_\gamma}=K_d\:e^{-E_\gamma/E_0^d}+K_t\:e^{-E_\gamma/E_0^t}
\end{equation}
The direct slopes of the three heaviest targets, $E_0^d
\approx$ 20 MeV, are two to three times larger than the
thermal ones, $E_0^t \approx$ 6 - 9 MeV, and the thermal
contribution represents 15\% - 20\% of the total yield (Table
\ref{tab:enterria:1}). No thermal component is apparent in the
photon spectrum measured in the $^{36}$Ar\/+\/$^{12}$C reaction and
direct bremsstrahlung alone accounts for the whole photon emission
already above $E_\gamma\approx$20 MeV (Fig. \ref{fig:enterria:1},
right). As a matter of fact, in such a light system there are not
enough nucleons in the overlap volume of target and projectile to
experience more than the minimal 2-3 collisions needed for
thermalization to take place (the average number of $pn$ collisions
in this system is $\langle N_{pn}\rangle$
= 1.8 according to the ``equal-participant" geometrical
model~\cite{enterria:Nife90}). Since no sufficient stopping and
equilibration are achieved, pure first-chance bremsstrahlung
dominates the entire hard-photon emission. Direct slopes,
independently of the system size, are proportional to the
laboratory projectile energy per nucleon, $\epsilon_{lab}$, as
expected for pre-equilibrium emission in prompt $NN$ collisions and
in good agreement with the systematics~\cite{enterria:Schu97}. This
dependence does not apply to thermal hard-photons: their slopes
scale with the total energy available in the nucleus-nucleus
center-of-mass, $\epsilon_{AA}$ (Fig. \ref{fig:enterria:2}),
strongly supporting the fact that thermal hard-photons originate at
later stages of the reaction after dissipation of the incident
energy into internal degrees of freedom over the whole system in
the $AA$ center-of-mass. The thermal slopes, lower by a factor two
to three than the direct ones, reflect the lesser energy available
in secondary $NN$ collisions.

\end{multicols}

\begin{table}
\squeezetable
\begin{center}
\caption{Direct and thermal slopes (inclusive, and at $\theta_{lab}$=90$^\circ\pm2^\circ$
to minimize Doppler effects), ratios of thermal to total
intensities, source velocities of both components, and total
cross-sections, for the hard-photons measured in
$^{36}$Ar\/+\/$^{197}$Au (inclusive, multifragment and central
reactions), $^{107}$Ag, $^{58}$Ni, $^{12}$C at 60{\it A} MeV. The
errors include statistical and systematic effects.
nss : not statistically significant.} 
\begin{tabular}{lccccccc} 
\small{System} &\small{$E_0^{d}$ (MeV)} &\small{$E_0^{t}$ (MeV)} &\small{$E_0^{t (90^\circ)}$ (MeV)} &\small{$I_{t}/I_{tot}$ (\%)} &\small{$\beta_{S}^d$} &\small{$\beta_{S}^t/\beta_{AA}$} & \small{$\sigma_{hard-\gamma}$ (mb)}\\ \hline
\small{$^{36}$Ar\/+\/$^{197}$Au} &\small{20.2 $\pm$ 1.2} &\small{6.2 $\pm$ 0.5} &\small{6.8 $\pm$ 0.6} &\small{19 $\pm$ 1} & \small{0.17 $\pm$ 0.02} & \small{1.6 $\pm$ 0.3} & \small{3.8 $\pm$ 0.2} \\ 
\small{$^{36}$Ar\/+\/$^{107}$Ag} &\small{20.1 $\pm$ 1.3} &\small{6.1 $\pm$ 0.6} &\small{7.0 $\pm$ 1.0} &\small{15 $\pm$ 1} & \small{0.18 $\pm$ 0.02} & \small{1.0 $\pm$ 0.3} & \small{3.1 $\pm$ 0.2} \\ 
\small{$^{36}$Ar\/+\/$^{58}$Ni}  &\small{20.9 $\pm$ 1.3} &\small{8.8 $\pm$ 0.8} &\small{9.0 $\pm$ 1.0} &\small{20 $\pm$ 1} & \small{0.18 $\pm$ 0.03} & \small{1.4 $\pm$ 0.3} & \small{1.8 $\pm$ 0.1} \\ 
\small{$^{36}$Ar\/+\/$^{12}$C}   &\small{18.1 $\pm$ 1.1} &\small{0.0 $\pm$ 0.5} &\small{0.0 $\pm$ 0.5} &\small{ 0 $\pm$ 2} & \small{0.20 $\pm$ 0.01} & \small{-} & \small{0.6 $\pm$ 0.1} \\ \hline 
\small{$^{36}$Ar\/+\/$^{197}$Au (multif.)} &\small{20.2 $\pm$ 1.3} &\small{6.0 $\pm$ 0.8} & \small{nss} &\small{16 $\pm$ 2} & \small{nss} & \small{nss} &\small{0.4 $\pm$ 0.1} \\ 
\small{$^{36}$Ar\/+\/$^{197}$Au (central)} &\small{19.7 $\pm$ 1.3} &\small{6.2 $\pm$ 0.8} & \small{nss} &\small{18 $\pm$ 2} & \small{nss} & \small{nss} &\small{0.4 $\pm$ 0.1} \\ 
\end{tabular}
\label{tab:enterria:1}
\end{center}
\end{table}

\begin{multicols}{2}

The thermal origin of the second hard-photon component is confirmed
by the analysis of the (Doppler-shifted) laboratory angular
distributions. The hard-photon angular distributions of the three
heaviest systems yield a source velocity which is systematically
lower (by a factor 15\%) than the $NN$ center-of-mass velocity,
$\beta_{NN}$, when assuming that they are emitted from a single
moving source \cite{enterria:Dente00b,enterria:Dente00}. They can
be well described, however, considering a midrapidity emission
($\beta_S^d\approx\beta_{NN}\approx$0.18) with slope $E_0^d$ plus
an isotropic emission with slope $E_0^t$ from a source moving with
a reduced velocity close to the $AA$ center-of-mass, $\beta_{AA}$,
of each system (Table \ref{tab:enterria:1}). The ratios of thermal
to direct intensities are fixed, in this double-source analysis, by
the ratios measured in the energy spectra. 
All other potentially conceivable mechanisms for photon production
above E$_{\gamma}$ = 30 MeV (e.g. statistical photons from the high
energy tail of Giant-Dipole Resonance (GDR) decays, coherent
nucleus-nucleus, cluster-nucleus or $pp\gamma$ bremsstrahlung)
investigated elsewhere \cite{enterria:Dente00b}, cannot
consistently explain the full set of data.

\begin{figure}
 \begin{center}
  \mbox{\epsfxsize=8.0cm \epsfbox{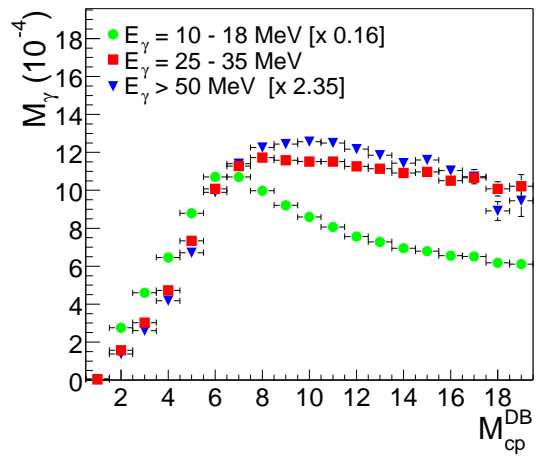}}
 \end{center}
 \caption{Photon multiplicity, M$_\gamma$, as a function of
the ``Dwarf-Ball" charged-particle multiplicity, M$_{cp}^{DB}$,
for: statistical nuclear photons (circles), thermal (squares) and
direct hard-photons (triangles), emitted in
$^{36}$Ar\/+\/$^{197}$Au at 60{\it A} MeV. M$_\gamma$[10-18 MeV]
and M$_\gamma$[$>$50 MeV] have been scaled to M$_\gamma$[25-35 MeV]
(the scaling factor is given in brackets). The M$_\gamma$
systematic errors (not shown) are below 10\%.}
\label{fig:enterria:3}
\end{figure}

The photon yield per nuclear reaction has been studied as a
function of impact parameter in the $^{36}$Ar\/+\/$^{197}$Au system
using the charged-particle multiplicity, $M_{cp}$, measured by the
phoswich multidetectors. We have classified the photons detected in
TAPS into three broad categories: (1) statistical nuclear decay
photons, with E$_{\gamma} =$ 10 - 18 MeV, (2) thermal
bremsstrahlung photons, in the range E$_{\gamma} =$ 25 - 35 MeV,
and (3) direct bremsstrahlung photons with E$_{\gamma}>$ 50 MeV.
Thermal and direct hard-photon multiplicities exhibit a very
similar dependence with charged-particle multiplicity (Fig.
\ref{fig:enterria:3}), increasing by a factor $\sim$10 from
$M_{cp}$ = 2 (peripheral) to $M_{cp}$ = 9 (semi-central
collisions). They stay roughly constant at $M_{\gamma}\approx$
1.2$\cdot$10$^{-3}$ for $M_{cp}$ = 9 - 18 (semi-central and central
reactions). This last class of events includes the reactions for
which the incident $^{36}$Ar is overlapped entirely by the much
larger $^{197}$Au target nucleus. Within this region of impact
parameters, the number of $NN$ collisions
saturates 
and so does bremsstrahlung photon production. On the contrary, the
statistical low-energy photon yield decreases for increasingly
central reactions by a factor $\sim$2 from $M_{cp}$ = 7 to
$M_{cp}\approx$20. This result is consistent with the quench of the
GDR photon yield (the dominant component in the region E$_{\gamma}
=$ 10 - 18 MeV \cite{enterria:Gaar92}) observed at high excitation
energies \cite{enterria:Gaar92,enterria:Suom98} and interpreted as
a result of the loss of collectivity due to a change from ordered
mean-field-driven motion to chaotic nucleonic motion.

\begin{figure}
 \begin{center}
  \mbox{\epsfxsize=7.5cm \epsfbox{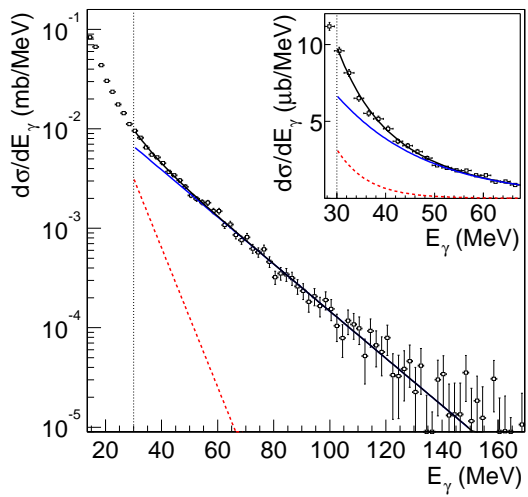}}
 \end{center}
 \caption{Hard-photon (E$_{\gamma}>$30 MeV) spectrum in
$^{36}$Ar\/+\/$^{197}$Au reactions with $M_{IMF}^{DB}\geqslant$3,
fitted to the sum of a direct (solid) and a thermal (dashed)
exponential. The thermal component is more apparent on a linear
scale in the inset.}
 \label{fig:enterria:4}
\end{figure}

Multi-fragment exit-channels, with $M_{IMF}^{DB}\geqslant$3,
account for 8\% of the total $^{36}$Ar\/+\/$^{197}$Au reaction
cross-section. Interestingly, those reactions (as well as the 9\%
most central reactions with $M_{cp}>$15) show also a thermal
bremsstrahlung component with roughly the same slope and intensity
as found in the inclusive case (Fig. \ref{fig:enterria:4} and Table
\ref{tab:enterria:1}).
The concurrent observation of IMF and thermal bremsstrahlung
emission in Ar+Au sets a lower limit for the fragmentation
time-scale. All microscopic transport models
\cite{enterria:Rich00,enterria:Cass90,enterria:Aich91}, in
agreement with the available data \cite{enterria:Dura99}, indicate
that the primary mechanism in intermediate-energy heavy-ion
reactions are quasi-binary dissipative processes in which two
excited and expanding quasi-target and quasi-projectile emerge
right after having traversed each other. The transient time up to
the moment when both ions loose contact is t
$\approx2R_{A}/\beta\approx$ 80 $\pm$ 20 fm/c, a time that
characterizes the end of the pure pre-equilibrium phase
\cite{enterria:Rich00}. If, as observed, thermal $NN\gamma$
collisions occur within the produced hot system, two conditions
have to be met: i) equilibration has been previously reached, and
ii) the radiating source has not yet fully fragmented. Microscopic
calculations of the Boltzmann-equation \cite{enterria:Cass90} or
molecular-dynamics \cite{enterria:Aich91} type for the
$^{36}$Ar\/+\/$^{197}$Au system
\cite{enterria:Mart95,enterria:Dente00b} show that second-chance
$NN\gamma$ collisions take place at t$\approx$ 100-200 fm/c during
the first compression undergone by the heavy quasi-target when
it reverts towards normal density after separating from the projectile. 
Thus, the source has to break-up at times, t$\gtrsim$150 fm/c, more
consistent with a ``sequential" scenario \cite{enterria:Frie99},
than with a fast (t$<$100 fm/c) fragmentation in a more dilute
state. The validity of such conclusions for central collisions of
more symmetric systems is presently under study
\cite{enterria:Orte00}.

In summary, hard-photon production has been investigated in four
Ar-induced reactions at 60{\it A} MeV. 
Aside from the dominant emission in first-chance (off-equilibrium)
$NN\gamma$ collisions, a second thermal bremsstrahlung component is
observed. The thermal origin of these photons is inferred based on:
i) their presence in the three heaviest systems,
accounting for 15\%-20\% of the total hard-photon yield, and their
absence in the light $^{36}$Ar\/+\/$^{12}$C system which does not
provide enough secondary collisions; ii) the scaling of the thermal
slopes with the $AA$ center-of-mass energy as expected for a
process taking place after dissipation of the energy into internal
degrees of freedom; iii) the consistency of the source velocity
analysis with a later isotropic emission for the thermal component;
iv) the significantly different behaviour of statistical GDR photon
and thermal hard-photon yields for increasing centrality suggesting
a change from nuclear collective to nucleonic chaotic motion. The
persistence of the thermal component in multifragment
$^{36}$Ar\/+\/$^{197}$Au reactions sets a longer time-scale for
nuclear multifragmentation than expected from pure dynamical or
simultaneous statistical models. 

We thank the AGOR accelerator team for providing a high-quality
beam. We gratefully acknowledge the loan of the ``Dwarf-Ball"
array and the support from W.U.~Schroeder, L.~Sobotka and
J.~T\~oke. This work has in part been supported by the IN2P3-CICYT
agreement, the Stichting FOM, and by the European Union HCM network
under Contract No. HRXCT94066.


\end{multicols}

\end{document}